\documentclass{infeduarxiv}
\usepackage{booktabs}
\usepackage{multirow}
\usepackage{xurl}

\volume{}
\issue{}
\pubyear{2025}
\articletype{research-article}
\doi{}

\begin{document}

\begin{frontmatter}

\pretitle{Research Article}

\title{The Hands-Up Problem and How to Deal With It: Secondary School Teachers' Experiences of Debugging in the Classroom}

\author{\inits{L.}\fnms{Laurie}
\snm{Gale}\ead[label=e1]{lpg28@cst.cam.ac.uk}\orcid{0009-0004-4299-6704}}

\author{\inits{S.}\fnms{Sue}
\snm{Sentance}\ead[label=e1]{ss2600@cst.cam.ac.uk}\orcid{0000-0002-0259-7408}}

\address{Raspberry Pi Computing Education Research Centre, \institution{University of Cambridge}, \cny{United Kingdom}}

\runtitle{The Hands-Up Problem and How to Deal With It}

\begin{abstract}
Debugging is a vital but challenging skill for beginner programmers to learn. It is also a difficult skill to teach. For secondary school teachers, who may lack time or programming experience, honing students’ understanding of debugging can be a daunting task. Despite this, little research has explored their perspectives of debugging. To this end, we investigated secondary teachers’ experiences of debugging in the classroom, with a focus on text-based programming. Through thematic analysis of nine semi-structured interviews, we identified a common reliance on the teacher for debugging support, embodied by many raised hands. We call this phenomenon the `hands-up problem'. While more experienced and confident teachers discussed strategies they use to counteract this, less confident teachers discussed the negative consequences of this problem. We recommend further research into debugging-specific pedagogical content knowledge and professional development to help less confident teachers develop approaches for supporting their students with debugging.
\end{abstract}


\end{frontmatter}

\section{Introduction}
The process of finding and fixing errors in a computer program is one of the most fundamental but challenging skills to learn for the beginner programmer. Learners must have sufficient understanding of several components of the programming language they are using to effectively and reliably resolve errors, which is rarely present in practice. Instead, beginners are commonly observed utilising ineffective debugging strategies (e.g., \citep{ConditionsLearningNovices, FindingFixingFlailing, ProblemSolvingDebuggingK68}) which can easily stall progress and evoke emotional distress \citep{ProgrammingAssignmentsEmotionalToll, EDAProgrammingEmotions}.

Secondary school teachers play a key role in developing their students' debugging ability. In an ideal classroom environment, they are on hand to teach students effective debugging strategies, support those struggling to debug, and foster positive attitudes towards errors. In practice, this is not always the case, with significant challenges unique to the school context. Teachers may not be confident in their programming ability \citep{ExpandingCSChallengesExperiences, RS2017Report, ComputingInCurriculumChallengesStrategies}, lack the lesson time to explicitly teach debugging \citep{ComputingInCurriculumChallengesStrategies, CurrentPerspectivesDebugging}, or have to support a large number of students with fixing errors \citep{CurrentPerspectivesDebugging}. These challenges are arguably exacerbated when teaching text-based programming languages (TBPLs), which is stipulated in many secondary-level national curricula (e.g., \citep{UKSecondaryComputingCurriculum, BritishColumbiaADSTCurriculum, PolandCSCurriculumChallengesSolutions, ProgrammingSkillsInLatinAmerica, Swedish2024Curriculum}). Although students perceive these languages as more realistic and expressive than block-based languages \citep{ToBlockOrNotToBlock}, they have a far larger scope for errors and subsequent struggle with resolving them \citep{JadudCompilationBehaviour, UnderstandingSyntaxBarrier}. Teaching with TBPLs also poses an additional layer of difficulty for time-poor or confidence-lacking teachers. These factors can easily make debugging a significant source of negative experiences for students and teachers in secondary school programming lessons.

While most debugging education research has focused on the student perspective, teachers' classroom experiences are far less studied. This work is always important to conduct; understanding teachers' experiences is essential for ensuring that research remains close to the challenges teachers are facing in the classroom. Focusing on debugging is particularly important as (text-based) programming becomes more widely taught in schools around the world.

This paper reports on secondary computing teachers' experiences with the teaching and learning of debugging in a text-based programming language in their classrooms. Through a thematic analysis of semi-structured interviews with nine teachers, we report on teachers' current approaches for teaching various aspects of debugging, their observations of students engaging in the debugging process, and their challenges with teaching it. The research questions for this study are:

\begin{itemize}
    \item RQ1: What are secondary school computing teachers' experiences and perspectives relating to debugging in the classroom?
    \item RQ2: What are secondary school computing teachers' observations relating to their students' debugging experiences?
\end{itemize}

Together with related work, we use our findings to conceptualise a significant in-class challenge for computing teachers that derives from student struggle with debugging, titled \textit{the hands-up problem}. We introduce this as a representation of a classroom environment unconducive to positive experiences with learning to program in Section \ref{sec:discussion}. We end with recommendations on how the hands-up problem could be alleviated and further developed as a concept. Our findings can be used by school teachers and researchers seeking to understand other teachers' classroom experiences with debugging and improve the classroom environment of programming lessons.

\section{Related Work}\label{sec:related-work}
An increasing amount of computing education research is being directed towards the teaching and learning of debugging. This is for good reason: debugging is a core aspect of programming, difficult to master, and has well-documented challenges. Even as generative artificial intelligence becomes more prevalent in programming education, debugging is still, and may be increasingly, a vital aspect of programming \citep{ComputingEducationGenAIEra}.

\subsection{Student Perspectives of Debugging}
Most debugging education research has focused on the student perspective. Students learning to program naturally have fragile knowledge \citep{FragileKnowledgeOriginalPaper} and misconceptions which often manifest as errors in their programs \citep{PeaConceptualBugs, CommonLogicErrors, MisconceptionsBeginnerProgrammer}. For students learning a TBPL, the scope for errors is naturally far larger than in BBPLs. Even minor syntactical errors can pose a significant barrier to a working program \citep{JadudCompilationBehaviour, UnderstandingSyntaxBarrier}, not helped by unclear programming error messages \citep{ErrorMessageWorkingGroup}. These challenges arguably make debugging in TBPLs more difficult than in BBPLs.

When trying to fix these errors, strategies that do not aid successful debugging have been observed among students using block-based \citep{ProblemSolvingDebuggingK68, ElementaryPuzzleBasedDebugging} and text-based \citep{FindingFixingFlailing, ThinkAloudNoviceDebugging} environments. These often embody a cycle of unreflective tinkering \citep{ConditionsLearningNovices}, which makes struggle with debugging likely. This struggle can evoke strong negative emotions \citep{ProgrammingAssignmentsEmotionalToll, EDAProgrammingEmotions} and ultimately dampen students' experiences of learning to program.

The challenges beginner programmers face when debugging have motivated an increased focus on debugging-specific teaching aids over the last few decades. These include systematic debugging processes (e.g., \citep{CarverImprovingChildrensDebugging, SystematicProcessMichaeli}), debugging-specific tooling (e.g., \citep{PythonTutor, Ladebug, SystematicDebuggingLogicalErrors}) and beginner-friendly programming error messages (e.g., \citep{EffectiveApproachErrorMessageEnhancement, ErrorMessageReadability, HighPrecisionFeedbackSyntaxErrors}). Many of these have been specifically designed for TBPLs. Summaries of pedagogical debugging `interventions' have found initial positive effects on measures of debugging ability \citep{DebuggingMetaAnalysis, DebuggingInterventionLitReview}, although evidence of wider adoption is lacking.

\subsection{Teacher Perspectives of Debugging in the Classroom}
Compared to student-centric research, little work has focused on teachers' perspectives on teaching debugging. Naturally, students' experiences with debugging in the classroom are mediated by the teacher, who may employ a range of strategies to develop students' confidence in resolving errors. However, school teachers face many challenges in providing high-quality debugging instruction. At a subject level, teachers have reported a lack of time and confidence to teach computing \citep{ExpandingCSChallengesExperiences, RS2017Report, ComputingInCurriculumChallengesStrategies}. As a consequence, teachers may lack time to explicitly teach debugging or be inundated with students in need of support \citep{CurrentPerspectivesDebugging}.

To our knowledge, very few studies have specifically investigated teachers' experiences of debugging in the classroom, despite the aforementioned challenges. Rather, related work has generally investigated teacher-student interactions or concepts related to pedagogical content knowledge.

\subsubsection{Teacher-Student Interactions}\label{sec:teacher-student-interaction}
In a classroom setting, teachers frequently assist students with debugging \citep{CurrentPerspectivesDebugging}. These interactions vary widely, but can generally be viewed through the lens of \textit{cognitive apprenticeship} \citep{CognitiveApprenticeshipOriginalArticle}, an instructional paradigm for teaching problem-solving processes using the principles of traditional apprenticeship.

Teachers' use of \textit{scaffolding} is a major theme in many studies of teacher-student interactions. A popular and general scaffolding technique is to ask students questions, with the content of the questions determining the level of scaffolding provided. Some teachers ask open-ended and generic questions \citep{SupportingStudentsScienceInquiry, ReflectionsSustainedDebuggingSupport, TeacherSupportForDebuggingPhysicalComputing}, providing relatively little scaffolding, while other questions may be far more specific \citep{MeasuringCSPCK, InteractionalWorkLearningDebug, SupportingStudentsScienceInquiry, TeacherSupportForDebuggingPhysicalComputing, TeacherDiagnosticInterventionSkills}. Teachers' initial questions to students may help the teacher to understand the error(s) in students' programs \citep{TeacherSupportForDebuggingPhysicalComputing, TeacherDiagnosticInterventionSkills}. More open-ended questions also encourage students to \textit{articulate} their thoughts, another aspect of cognitive apprenticeship important for debugging.

More debugging-specific techniques include \textit{modelling} the debugging process \citep{ReflectionsSustainedDebuggingSupport, TeacherDiagnosticInterventionSkills}. Unsurprisingly, students' debugging strategies sometimes differ from their teachers', at which point teachers may decide to tailor their approach to the students' strategies \citep{ReflectionsSustainedDebuggingSupport}.

While general techniques used by teachers in their interactions with students are useful to know, the support provided is incredibly contextual. \citet{SupportingStudentsScienceInquiry} frame teacher-student interactions as improvisational, meaning teachers' general approaches must be adapted to a specific interaction. Other in-class factors, such as the time remaining in the lesson, the number of students in the classroom, and the type of error in question, also influence the support the teacher will provide \citep{TeacherSupportForDebuggingPhysicalComputing}. In the same vein, these studies each have their own context, some of which are not representative of an actual classroom environment. Some were conducted in summer schools \citep{ReflectionsSustainedDebuggingSupport} or simulations of classrooms \citep{SupportingStudentsScienceInquiry, TeacherDiagnosticInterventionSkills}, so not all findings necessarily reflect teachers' actual in-class interactions with students.

\subsubsection{Pedagogical Content Knowledge and Professional Development}
The approaches that teachers adopt when providing debugging support, and the knowledge required to employ them, form part of their `debugging pedagogical content knowledge' (PCK) \citep{PCKOriginalArticle, RefinedConsensusModel}. \citet{PCKOriginalArticle} defines this as ``the ways of representing and formulating the subject that make it comprehensible to others'' (p. 9). PCK can be developed in several ways: teaching experience, research-informed teaching resources, or professional development (PD) \citep{PCKOriginalArticle}. Some debugging-specific PD activities have been developed in previous work, involving frequent teacher reflection \citep{EffectsOfProfessionalDevelopmentK12Debugging} and simulating the student perspective \citep{DoINeedToKnowIfTheTeacher, SupportingStudentsScienceInquiry}. Although analyses of these activities only included small cohorts of teachers, they reported benefitting from a low-stakes environment to improve their knowledge of debugging strategies, common errors, and teaching approaches \citep{DoINeedToKnowIfTheTeacher, EffectsOfProfessionalDevelopmentK12Debugging, SupportingStudentsScienceInquiry}.

Despite the volume of pedagogical and content knowledge required to teach debugging effectively, we are not aware of a model of PCK specifically for debugging.  One study analysed school teachers' level of debugging PCK \citep{DebuggingPCK}, but did not propose a formal model for it and focused on Scratch. Related work has focused on developing computing or programming PCK more generally \citep{ProgrammingPCK, CSPCKConceptualisation}. A related concept that could inform the development of a debugging PCK is the idea of debugging learning trajectories \citep{DebuggingLearningTrajectory, DebuggingLearningTrajectoryTextBasedLanguages}. Although designed for curriculum development, they could be used to intuitively visualise content knowledge to teachers. This would be particularly helpful for non-specialist teachers wanting to understand fundamental debugging concepts.

Overall, relatively little research has been conducted into teachers' experiences of debugging in the classroom. Studies on teacher-student debugging interactions indicate that some teachers are aware of a variety of teaching approaches that can be applied to debugging. However, several of these studies do not take place in the classroom, where teachers' time for interactions with students may be severely limited \citep{CurrentPerspectivesDebugging}. For less experienced or confident teachers, debugging-specific PD is a helpful way of developing teachers' debugging PCK, which itself is a concept that could be further developed.

\section{Method}
To explore both teachers' experiences and perspectives with teaching and supporting students with debugging in the classroom, we conducted semi-structured interviews with nine teachers.

\subsection{Participants}
Teachers were purposively sampled for the study, with the criteria that they must be teaching a text-based programming language at a lower secondary level\footnote{By lower secondary, we refer to the first three years of secondary education, known as key stage three in England, equivalent to grades 6-8 and ages 11-14.}. Eligible teachers were invited to partake in the study through a teacher research network newsletter, hence incorporating an element of convenience sampling.

Nine teachers took part in the study. Seven of these were from England, where the average class size for state-funded schools is 26 students \citep{UKAverageClassSizes}, and two were from other countries. Each of these teachers\footnote{Apart from one teacher, who was between teaching jobs at the time of interviewing.} taught computing as a compulsory subject to their lower secondary students. Teachers from state-funded and non-English schools followed the English computing curriculum, which mandates the ``use two or more programming languages, at least one of which is textual''  \citep[p. 2]{UKSecondaryComputingCurriculum}. Teachers from private or independent schools were able to implement their own programme of study. Python was taught by all teachers, with some also teaching other TBPLs such as Swift. Prior to learning these TBPLs, their students would have been taught a BBPL in primary school \citep{UKPrimaryComputingCurriculum}. Each teacher received a bookshop voucher worth \pounds10 as a token of appreciation for participating. Table \ref{tab:demographic-breakdown} reports the demographic information of the teachers interviewed.

\begin{table}
 \caption{Demographic Information For Participating Teachers}
 \label{tab:demographic-breakdown}
    \begin{tabular}{p{3.5cm} p{8cm}}
        \toprule
         \textbf{Characteristic} & \textbf{Demographic Breakdown} \\
         \midrule
         Gender & Male: 4 \hspace{0.5cm} Female: 5 \\
         Age & 30-39: 2 \hspace{0.5cm} 40-49: 3 \hspace{0.5cm} 50-59: 3 \\
         Years of computing teaching experience & 0-5: 1 \hspace{0.5cm} 6-10: 2 \hspace{0.5cm} 11-15: 3 \hspace{0.5cm} 16-20: 2 \hspace{0.5cm} Over 20 years: 1 \\
         School information & Mixed-gender: 8 \hspace{0.5cm} State-funded: 5 \hspace{0.5cm} Private or independent: 2 \hspace{0.5cm} Grammar: 2 \\
         \bottomrule
    \end{tabular}
\end{table}

\subsection{Procedure and Data Collection}
We conducted semi-structured interviews to allow teachers ample time and flexibility to discuss matters relevant to their experiences of debugging in the classroom. Moreover, as per our research questions, we were particularly interested in teachers' \textit{reported} feelings and experiences, which suited interviews well.

Interviews took place virtually from October to December 2023, using Google Meet or Microsoft Teams. Each teacher was interviewed individually, apart from Chloe and Deborah, who were interviewed together. Before being interviewed, each teacher completed a consent form and demographic questionnaire, in accordance with the study procedure approved by our department's ethics committee. Interviews lasted between 35-65 minutes.

The interviews facilitated two main points of discussion. Firstly, teachers' \textit{own experiences and perspectives} with debugging in the classroom. That is, their personal thoughts and recollections related to debugging in programming lessons. This could include teaching approaches they used, general feelings they have experienced about debugging, or beliefs around the role of debugging when teaching programming. Second, we asked teachers about their \textit{observations of students' experiences} with debugging, including any common errors, behaviours, or reactions that they commonly see among students. The interview schedule, along with the codebook definitions and agreement procedure, is available to view in the study repository \citep{StudyRepository}. We did not have consent from the teachers to publish the transcripts.

Each interview was recorded and automatically transcribed. The transcripts were then manually read through in conjunction with the recording to correct any mistakes and remove any mentions of personal or school-specific information, while also being used for initial memoing. These transcripts were sent to teachers within two weeks of the interview, who were given two weeks to mention anything they wished to be omitted or corrected. For the teachers who took this opportunity, a final version of the transcript was sent to them for their records. After this period, transcripts were disassociated with the teachers' contact details.

\subsection{Analysis}
We analysed teachers' interviews using `codebook' thematic analysis (TA) \citep{ThematicAnalysisOriginalArticle, ConceptualDesignThinkingThematicAnalysis}. This is a `school' of TA that maintains the core values of reflexive TA, in which the subjectivity and experiences of the researcher are key aspects of the analysis, while incorporating some pragmatic approaches that seek reproducible coding \citep{ConceptualDesignThinkingThematicAnalysis}. In practice, we conceptualised latent themes that identified meaning beyond teachers' responses, as discussed in \citet{ReflectingReflexiveThematicAnalysis, GoodPracticeThematicAnalysis}, while incorporating approaches to ensure the codebook was consistently applied across the interview data (see Section \ref{sec:reliability-validity}).

This began with multiple passes of the interview transcripts. The first of these, as previously mentioned, was used both for correcting the automatically generated transcripts and initial memoing. This was performed by the first author. The second author then read through the corrected transcripts and wrote initial memos.

After this, the interview transcripts were migrated to NVivo 12, which was used for the rest of the analysis. Each transcript was then read through again by the first author to inductively generate open descriptive codes and note further memos. Through further inspection of the interview data, open codes were inductively abstracted into higher-level latent themes and sub-themes that aimed to provide an interpretive, and hence subjective, account of the data. This process of theme generation involved repeatedly restructuring the codebook, redefining the themes and sub-themes, and rewriting the report of the codebook, encompassing steps 4-6 in \citet{ThematicAnalysisOriginalArticle}. This process was conducted by the first author with feedback from the second author after some of the iterations. Our approach was therefore more iterative than linear, which is considered a normal part of TA \citep{ThematicAnalysisOriginalArticle}.

The analysis culminated in a codebook containing six themes, each containing sub-themes, some of which contained their own sub-themes. Only the lowest-level sub-themes were used to code the data. The final commentary of the results aims to provide a rich description of all portions of the interviews that are relevant to our research questions.

Regarding positionality, the first author is a doctoral student with limited previous experience in conducting interviews and qualitative analysis. The regular feedback of the second author was hence important, who is an experienced computing education researcher with expertise in qualitative analysis.

\subsection{Reliability and Validity}\label{sec:reliability-validity}
The content validity of the interview questions was strengthened through two procedures. First, the questions were iteratively refined by both authors to ensure they appropriately discussed teachers' experiences, perspectives, and observations of debugging. Second, a pilot interview was conducted with a teacher to assess the wording of the questions and the timings of the interview. This teacher was personally contacted by the first author and received the same bookshop voucher as the other participating teachers. Their transcript was not included in the analysis.

The reliability with which the codebook was applied was improved by seeking agreement between the authors' coding patterns without calculating an interrater reliability (IRR) coefficient. \citet{UnitisationProblemSemiStructuredInterviews} call this \textit{negotiated agreement}. We wanted to verify the consistency with which the first author coded the transcripts according to the codebook definitions, also known as \textit{intracoder reliability} \citep{IntercoderReliabilityQualAnalysisGuidelines}. Seeking agreement by identifying disagreements in our coding was appropriate for this case \citep{UnitisationProblemSemiStructuredInterviews, IRRHCI}. This was particularly important as the first author was inexperienced in qualitative analysis. Negotiated agreement hence improved the clarity of definitions and the consistency with which the first author coded.

Calculating an IRR coefficient was not required for the above process and was considered detrimental for our study. Our interview transcripts were long and conversational, including frequent tangents and interruptions. Naturally, this makes it difficult for coders to agree on units of meaning, which is required for calculating agreement with IRR coefficients. Combined with the complexity of the codebook, any IRR coefficient is likely to be positively skewed and unrepresentative of the reliability of the codebook \citep{UnitisationProblemSemiStructuredInterviews}. Furthermore, there is little consensus on acceptable IRR coefficient values for semi-structured interview data, which undoubtedly differs from other qualitative data with different research objectives \citep{ReliabilityQualAnalysisMisconceptions, UnitisationProblemSemiStructuredInterviews}.

Agreement was hence sought as follows. As described in the previous section, the first author developed the codebook with regular input from the second author. Two interview transcripts (Gideon and Hannah) were then randomly selected to be coded by both authors, in line with the guidelines of \citet{IntercoderReliabilityQualAnalysisGuidelines}. For this, the second author was provided with the codebook, complete with definitions, and an uncoded version of the transcripts. Once we both coded the transcripts, disagreements in coding were identified using the coding stripes feature in NVivo. Disagreements and ambiguous sub-themes were then discussed. Where necessary, revisions to the structure and definitions of the codebook were made.

\subsection{Limitations}
A limitation of our method was that we kept the final themes close to the research questions (i.e., closely related to debugging). This was done to avoid generating an overly large codebook, but in the process likely missed some information that contributes to teachers' experiences of debugging in the classroom.

Additionally, our sample was, within certain constraints, self-selecting. Although we were not seeking a generalisable population for an in-depth qualitative study, the participating teachers were likely engaged with computing education research to some degree. As a result, they may be more reflective or intentional about their teaching practice than most teachers, potentially affecting the themes we identified.

\section{Results}
Six themes related to teachers' experiences, perspectives, and observations of students were yielded from the interviews, which are:

\begin{itemize}
    \item Barriers to successful debugging
    \item The emotional nature of debugging
    \item Reliance on the teacher
    \item Varying levels of scaffolding for helping students to debug
    \item Improving students' debugging ability
    \item Promoting a positive error culture
\end{itemize}

The codebook structure and code definitions are available to view in the study repository \citep{StudyRepository}. We now report each of the themes, with sub-themes reported in italics, supported by teachers' quotes\footnote{All teachers have been given pseudonyms and all quotes have had fillers and contractions removed.} from the interviews.

\subsection{Barriers to successful debugging}
A particularly widespread theme was the obstacles that impede students from successfully debugging their programs. These challenges related to schools' allotted time for teaching programming, students' attitudes, and the nature of the Python language.

A barrier outside of the teachers' control was the \textit{lack of time for sufficient debugging practice}. Teachers were generally dissatisfied with the lesson time they had to teach programming, sometimes constrained to as little as six hours per year. As well as making any progress beyond basic programming constructs difficult, this has effects for debugging. Teachers simply do not have the lesson time for their students to frequently encounter common errors, nor teach effective debugging strategies. As succinctly put by Adam:

\begin{quote}
    \textit{``Do I have frustrations teaching students how to debug? Yeah, we don't have enough time to do it!''} - Adam
\end{quote}

Barriers related to the difficulties of learning to program (\textit{programming-related}), some of which were specific to learning a TBPL, also make debugging difficult for students. The \textit{difficulty of getting syntax exactly right} and \textit{conceptual difficulties}, sometimes manifesting as logical errors, were both mentioned as struggles for students. Many teachers described the common errors students encounter, mostly related to minor incorrect formatting, illustrating the `syntax barrier' \citep{UnderstandingSyntaxBarrier} present for many beginner programmers. Esther makes the point that such syntax errors may not be minor `slips of the hand', but a \textit{``a lack of the deeper understanding of what the language is actually trying to do''}.

Some teachers also commented on the \textit{hard to interpret programming error messages} in the Python environment students were using\footnote{At the time of the interviews, the most recent stable version of Python was 3.12.}. While they can be positively utilised (see Section \ref{sec:improving-debugging-ability}), teachers expressed the unclear and intimidating nature of them for students, not helped by the red text they usually appear in. Hannah explains that the content of the error message is not pitched at a level that most students understand, in turn increasing the difficulty of resolving syntax or runtime errors.

\begin{quote}
  \textit{``The error message that comes back is really quite cluttered \dots It could just be, `here's the line number'. And it's not very in student speak, so it will say like, `identification error'.''} - Hannah
\end{quote}

Not all the barriers to successful debugging identified by teachers were external to the student. Students may also have \textit{unhelpful attitudes towards computing or programming} which impede their ability or motivation to debug.

\begin{quote}
    \textit{``Why don't they like to debug? It's because \dots they don't like coding, they just don't like the lesson, they don't like the subject \dots they've never known the success because they've given up too early or they've been put off it.''} - Gideon
\end{quote}

Gideon highlights that students' negative attitudes span from debugging as a skill to computing as a subject. He also alludes to a lack of success and motivation as drivers for these negative attitudes. The former relates to the `performance accomplishment' determinant of self-efficacy \citep{OriginalSelfEfficacyArticle}, which is derived from a learner's initial experiences with a subject. For most students in England, initial experiences of programming come at primary school in block-based environments. However, unhelpful attitudes could also arise from students' experiences of learning more generally, which Isabel believes to be the case.

\begin{quote}
    \textit{``Maybe some of them do not expect themselves to finish successfully because \dots of their previous experiences in, not just in computing, but across the board of other subjects.''} - Isabel
\end{quote}

The effect of these attitudes often manifests as a lack of motivation to debug their programs. This has consequences for the teacher, who has the challenge of encouraging their students to persevere with debugging.

\subsection{The emotional nature of debugging}
In addition to attitudes, teachers discussed a range of emotional reactions they observe when their students are debugging. Observed reactions were generally physical or verbal, occurring at the point of encountering an error or after attempting to resolve an error. Visible reactions are not universal, however. Some teachers specifically pointed out the \textit{reserved reactions of quieter students}.

Most observed reactions represented \textit{negative emotion during debugging struggle}, conveying feelings such as frustration and disappointment. An interesting observation mentioned by several teachers is the emotional attachment students develop towards their programs and the distress this can cause when errors occur. For students, encountering errors in their own programs can be far more costly than debugging an erroneous program written by the teacher, leading to more dire emotional reactions. Felix describes this as potentially depressing for students.

\begin{quote}
  \textit{``When they’re starting to write upwards of ten lines of code, you're invested in getting it right and then if it fails, that can flip and that can be quite depressing.''} - Felix
\end{quote}

Repeatedly experiencing such negative emotions when debugging is difficult for students to deal with. Some students may therefore disengage from the debugging process if they become stuck too often.

\begin{quote}
    \textit{``Once they've hit the brick wall, that's kind of the green light to go, `I’m just gonna do anything'. [They] Sit there in silence, they’re not causing any commotion.''} - Gideon
\end{quote}

Despite the lack of visible frustration, disengagement may indicate more severe emotional distress, as students may not be able to cope with any more errors in their programs. Alternatively, students may disengage due to the difficulty of the error. Either way, some students' motivation to debug is reduced after repeatedly encountering errors. It can be difficult for busy teachers to notice these students who, as Gideon mentions, are \textit{``not causing any commotion''}.

Not all the emotional reactions reported were negative. The \textit{elation when solving an error} was mentioned by some teachers. These reactions were also acute, indicating feelings of intense joy and pride.

\begin{quote}
   \textit{``That dopamine hit when you get it right \dots I’ve seen kids struggling and struggling and struggling and all of a sudden, the light goes on or something happens \dots they even shake with excitement.''} - Felix 
\end{quote}

Felix's observations allude to the physical enactment of students' emotions, demonstrating the sheer joy that can be experienced from debugging. However, such positive reactions were rarely mentioned, perhaps due to the perseverance needed to experience them.

We also found some teachers alluded to the \textit{resilience of motivated students}, who do not experience such polarising emotions when debugging. These students remain composed and explore different solutions, not afraid to add errors to their programs. Hannah explains that most of her students have such resilience.

\begin{quote}
    \textit{``They are all interested in finding a solution to a problem. And I think that that's really important in having the resilience to bother to debug the program.''} - Hannah
\end{quote}

The resilience to debug, to a large extent, comes from the student. From Hannah's perspective, this derives from students' interest in programming. Other factors are undoubtedly at play, such as the debugging-specific teaching strategies teachers use (see Section \ref{sec:improving-debugging-ability}), their work in normalising errors in the classroom (see Section \ref{sec:positive-error-culture}), and students' access to computers at home. Hannah teaches in a highly selective school, where high levels of interest are perhaps not as surprising. Generally, however, negative emotions and disengagement were the most widely discussed observations we identified in this theme.

\subsection{Reliance on the teacher}\label{sec:reliance-on-teacher}
A combination of the \textit{barriers to successful debugging} and \textit{the emotional nature of debugging} contributes to many students relying on the teacher for debugging support. This is typically embodied by students raising their hands when encountering errors. Often, there are enough hands up to make it unrealistic for the teacher to effectively help every student requesting support.

Several teachers discussed \textit{the commonness of teacher reliance}. When asked about the atmosphere of her programming lessons, Esther simply stated \textit{``chaotic \dots lots of hands up and down all the time''} and that her students' problems are \textit{``almost always about errors in code''}. Chloe similarly reports \textit{``I think at least half the class will ask all the time''}; far beyond any teacher's capacity to meaningfully assist every student in classes of 25-30 students \citep{ExpandingCSChallengesExperiences, UKAverageClassSizes}.

Several teachers confidently posited that the cause of such reliance is a lack of student resilience or motivation, rather than the difficulty of the error they are debugging. Some students want their errors to be instantly resolved, which they cannot always do themselves. This creates a preference for waiting for teacher support over sustained independent effort.

\begin{quote}
    \textit{``They want you to give them the answer (laughs). They don’t want to think too much. Only very few of them would want to think and figure out the errors.''} - Deborah
\end{quote}

Once a student raises their hand, teachers generally stated that students do not continue debugging until their teacher approaches them. This is naturally detrimental for the student, with emotional reactions likely to ensue. However, this reliance also has a \textit{toll on the teacher}. Some teachers reported that \textit{``programming lessons are really, really busy lessons''} (Chloe), which can \textit{``erode their [teachers'] professional confidence''} (Adam). Esther talks about the obligation they feel to help every student, while Gideon discusses the challenge of not being able to do so.

\begin{quote}
    \textit{``You're always thinking about, `while their hands are up, they're not doing anything or making progress'. So you sort of feel this pressure to go and help everybody.''} - Esther
\end{quote}

\begin{quote}
    \textit{``This is the problem with debugging I suppose in general \dots I'll have my program and then I got 20 other programs, so I'm potentially debugging 25 programs in a lesson \dots This is beyond the capability of the human.''} - Gideon
\end{quote}

The impact of this pressure on the teacher is multifaceted. As well as the \textit{``pressure to go and help everybody''}, many teachers also feel a pressure to prove that their students have learnt something every lesson. This is very difficult when so many students' progress depends on the teacher.

Some teachers accept that it is not possible to personally help every student in need of debugging support. Instead, teacher reliance must be carefully managed. As Adam states, \textit{``there has to be a more scalable model than going around and solving individual people's code problems''}. Consequently, some teachers utilise strategies that \textit{motivate teacher-independent debugging}. These aim to provide students with the confidence to debug without the teacher by their side. For example, some teachers tell their students to consult other sources of information before approaching them. These may be their peers (\textit{encouraging helpful collaboration}) or resources that teachers have curated (\textit{signposting to resources for assistance}). The former does not just reduce the in-class burden on the teacher; explanations from peers \textit{``just seem to go in better''} (Hannah). Alternatively, some teachers will simply \textit{refuse to help students} until they have taken some effort to debug themselves, such as Hannah.

\begin{quote}
    \textit{``If they ask me any questions and they haven't tried to solve it themselves, I will just tell them, `you need to try it yourself first'. So there's not like an easy solution of, `I'm lazy. I'm just gonna ask for the answer'.''} - Hannah
\end{quote}

As a result, the teachers' role regarding debugging support becomes less diagnostic. In Isabel's words, \textit{``they do ask [for debugging support], we just redirect them''}. Students exposed to these strategies become accustomed to \textit{teacher-independent} debugging instead of \textit{teacher-dependent} debugging. Teachers who employ such approaches explained how this helps to develop resilience among students that is often lacking at the beginning of their programming journey.

\begin{quote}
    \textit{`They get better at these things, but there's just as many older students who make the same mistakes and they’re like  `oh, that was a silly error' rather than `that was a killer'.'' - Benjamin}
\end{quote}

\subsection{Varying levels of scaffolding for helping students to debug}
There are of course times when teachers do need to provide individualised debugging support to students. This initiates a teacher-student interaction, where teachers mentioned using a range of scaffolding techniques. Some of these are generic teaching tactics applied to the context of debugging, while others are programming- or debugging-specific. The level of scaffolding varies across a continuum, from very little, placing the onus on the student, to near-complete scaffolding of the problem, placing the onus on the teacher. The amount a teacher applies to a particular interaction partially depends on the teacher's knowledge of their student(s), as well as practical factors such as remaining lesson time \citep{TeacherSupportForDebuggingPhysicalComputing}.

When putting the onus on the student, some teachers simply ask students to share their thoughts about the error in question (\textit{encouraging externalisation of ideas} or \textit{asking students questions}). Both of these approaches encourage students to outwardly reflect on their understanding of the problem, which might be what is stopping them from successfully resolving errors.

\begin{quote}
    \textit{``It's that rubber ducking process, sometimes students don't explain it to themselves, and when they go back and explain their problem they spot it themselves.''} - Benjamin
\end{quote}

The questions teachers mentioned asking students were naturally of varying specificity, which again shifts the level of scaffolding. Adam deliberately uses generic questions at the beginning of interactions with his students.

\begin{quote}
    \textit{``You're gonna have to tell me what do you want it to do, what does it do at the moment and what have you tried already?''} - Adam
\end{quote}

These questions are useful for specialist and non-specialist teachers alike, as they do not require expert knowledge or time spent comprehending students' code. Furthermore, once students understand that they will be asked generic questions, they may be more inclined to debug independently of the teacher.

Heavier forms of scaffolding may be used if a student has been debugging for a sustained period of time, or if the error is difficult to localise. Some teachers may \textit{give students the solution} or \textit{give students hints} specific to the error(s) the student is struggling with.

\begin{quote}
    \textit{``If they're struggling to find it, I can see they're getting frustrated, first thing I would say is, `have a look at that particular line'. This is assuming I know where the error is.''} - Benjamin
\end{quote}

\subsection{Improving students' debugging ability}\label{sec:improving-debugging-ability}
As well as general scaffolding, most teachers reported using approaches designed to help students become more successful debuggers. A more general sub-theme of this theme is teaching \textit{programming strategies to make debugging easier}. These are patterns of programming that prevent errors from accumulating or help to keep a student aware of the purpose of their program. Adam and Hannah, for example, encouraged an iterative process of small changes and testing.

\begin{quote}
    \textit{``Every five minutes, they need a reminder, `you need to save and you need to debug'. And so that process of make one change, test it, make another change, test that, is really, really important.''} - Hannah
\end{quote}

Additionally, some teachers \textit{guide students through debugging strategies}. This typically involves explicitly modelling the debugging process with an example or student program. Live coding was a popular medium for this. Some teachers, such as Gideon, model to groups of students who are struggling with a particular error, in turn helping several students at once.

\begin{quote}
   \textit{``Sometimes if I'm getting two or three students with the same error, then I'll be like, `right come round the board. I need to demonstrate, I need to model'.''} - Gideon
\end{quote}

A few teachers were strong advocates of \textit{utilising programming error messages} (PEMs). Although they are not always intuitively phrased, Adam and Felix viewed PEMs as valuable sources of immediate feedback.

\begin{quote}
    \textit{``In Python, although it might look intimidating, that red line of text, actually that's really helpful because it's instant and it tells you where to start looking.''} - Adam
\end{quote}

This positive outlook on PEMs drives these teachers to frequently discuss them. Some teachers also mentioned teaching students how to `decode' PEMs.

Through repeated exposure and practice, guided by the strategies they teach, teachers reported a positive change in their students' debugging behaviours. Benjamin's students, who are \textit{``pre-cued to look for common errors''}, become drilled in doing this themselves.

\begin{quote}
    \textit{``They get better at looking for the obvious errors \dots the top five syntax errors, they will look for those straight away as they have the practice.''} - Benjamin
\end{quote}

This positive change appears to be more frequently reported by the more confident or experienced teachers we interviewed. Regardless, such improvement is to be celebrated. These teachers show how their teaching strategies help students to go about debugging with independence and confidence. With this comes more of the aforementioned \textit{elation of solving an error}, as well as positive initial experiences for students learning to program.

\subsection{Promoting a positive error culture}\label{sec:positive-error-culture}
Students' improvement in debugging is not solely due to instructional strategies. Cultivating a positive error culture, an environment where students feel comfortable making errors in their programs, also helps students become more resilient when debugging. One useful method for promoting a positive error culture is \textit{getting used to erroneous code}. This may involve sabotaging activities (Benjamin and Hannah), where students actively add errors to their peers' programs, or giving students erroneous code (Chloe and Gideon). Hannah points out that sabotaging is useful because it exposes students to errors in a depersonalised format, reducing the potential emotional angst around debugging.

\begin{quote}
    \textit{```It doesn't matter if it doesn't work. It's not yours. And so that was a really nice way to kind of break down the anxiety about that `there will be errors'.''} - Hannah
\end{quote}


Another method of normalising errors in the classroom is by highlighting that errors are to be expected rather than dreaded. Felix explains how `failures' should not just be reported, but celebrated to remove the stigma around them.

\begin{quote}
    \textit{``Spectacular failures. I want those reported and celebrated as well. If something should have gone right and [went] badly wrong but somebody found something interesting on the way \dots celebrate it. Take the fear out of it.''} - Felix
\end{quote}

Felix's response also highlights the practice of \textit{celebrating the effort as well as the outcome}. If students are actively rewarded for their sustained debugging efforts, as in Adam's case, this can act as a further motivator for teacher-independent debugging.

\begin{quote}
    \textit{``If we have a culture in a classroom where students know, `ok, sir’s not just going to give me a reward if I get things right \dots he’s gonna give me a reward if I try and get it right' and that's the big breakthrough for debugging.''} - Adam
\end{quote}

These approaches, combined with \textit{verbal encouragement} throughout the debugging process, help to create an environment where both encountering and struggling with errors are natural experiences. Such measures help make debugging a healthy, normal activity that can be a source of joy rather than despair.

\section{Discussion}\label{sec:discussion}
Teachers' experiences of debugging in the classroom spanned a range of insights and issues. A major motif we found among these experiences was the common reliance students have on their teacher for debugging support. This tended to be the case for less motivated students, who would raise their hands at the point of or soon after encountering an error.

Our findings, along with previous reports of teachers' experiences with debugging in the classroom, led us to conceptualise \textit{the hands-up problem}, the in-class difficulty deriving from the common student tendency to raise their hand when struggling to complete a task. Although our definition frames the hands-up problem as a potential challenge for teachers of any subject, we focus on programming classrooms due to the lack of similar studies with other subject teachers. Therefore, in this paper, `struggling to complete a task' refers to struggling to debug when learning to program.

We now introduce the hands-up problem as a challenge that computing teachers face in their programming lessons as a result of students struggling with debugging. We also propose ways to alleviate the problem. In addition to prior work, we reference theme names from the results section in italics where appropriate.

\subsection{The Hands-Up Problem}
The hands-up problem is the in-class difficulty deriving from the common student tendency to raise their hand when struggling to complete a task. Since the need for teacher support is reportedly common (see \textit{the commonness of teacher reliance} and \citet{CombattingLearnedHelplessness, CurrentPerspectivesDebugging, LackPerserveranceAutonomyDebugging, StumpTheTeacher}), the problem derives from many students raising their hands at once. Therefore, the severity of the problem scales with the size of the classroom, making it difficult for school teachers whose classes contain 25-30 students \citep{ExpandingCSChallengesExperiences, UKAverageClassSizes}.

\begin{figure}
    \centering
    \includegraphics[width=0.85\linewidth]{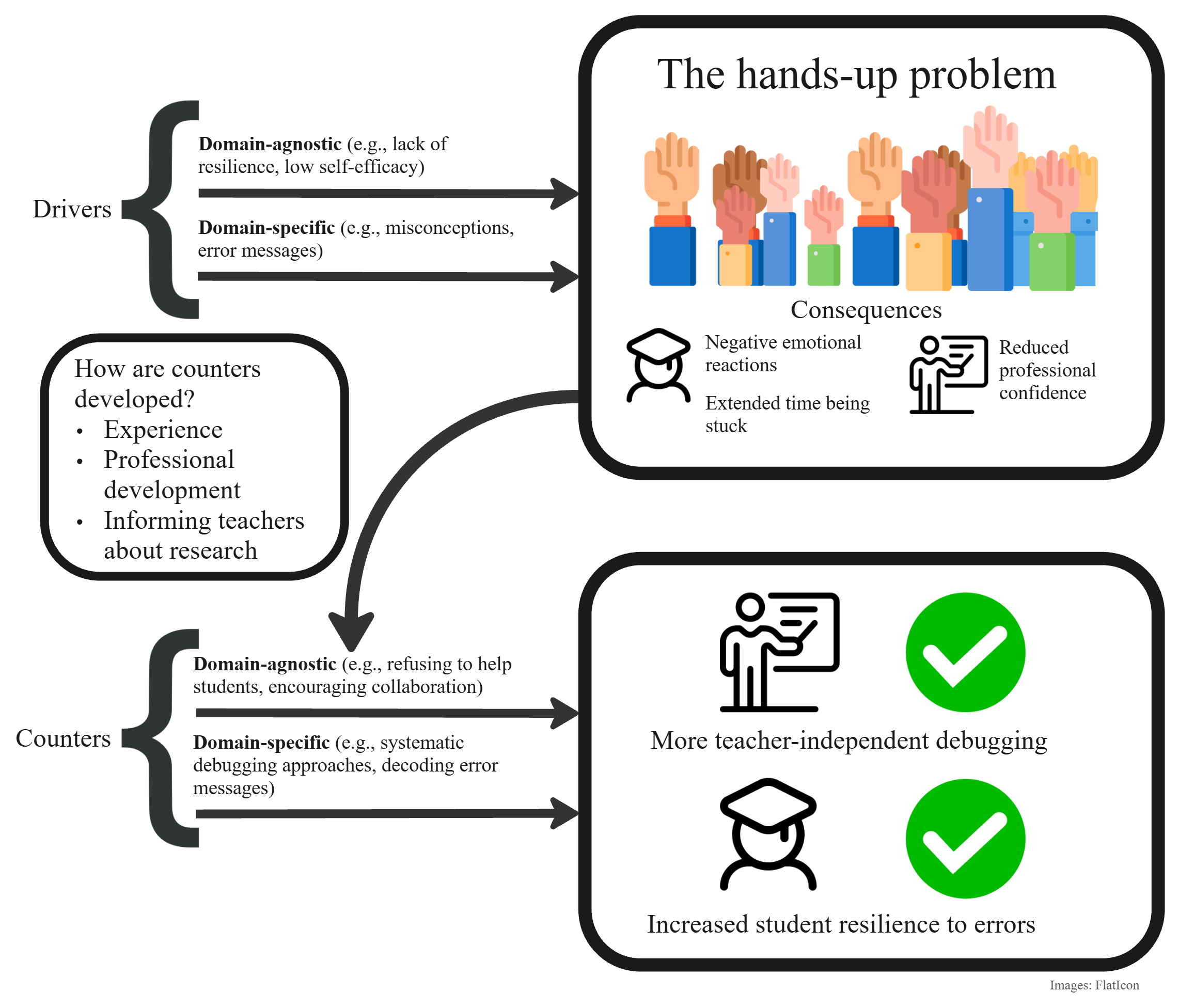}
    \caption{A Visualisation of the Hands-Up Problem}
    \label{fig:hands-up-problem}
\end{figure}

According to the teachers we interviewed, students generally raised their hands at the point of encountering an error. Hand-raising may also occur after students have put some initial effort into debugging \citep{CurrentPerspectivesDebugging, TeacherDiagnosticInterventionSkills}. Either way, the consequences for teachers and students are generally negative. We now describe some drivers, consequences, and counters to the hands-up problem, visualised in Figure \ref{fig:hands-up-problem}.

\subsubsection{The Drivers}
There is a range of \textit{drivers} that prompt a student to raise their hands during debugging struggle. These may be \textit{domain-agnostic} or \textit{domain-specific}. The list below is not meant to be exhaustive, but gives a general categorisation of some factors at play.

\textit{Domain-agnostic} drivers for hand-raising are factors that can contribute to the hands-up problem in any subject. We argue that these are generally related to students' affect. Teachers in our study often discussed a general lack of resilience to work through a problem, leading to reliance on the teacher. This problem is also highlighted by teachers in related work \citep{ComputingInCurriculumChallengesStrategies, TeacherDiagnosticInterventionSkills} and likely relates to students' motivation levels. However, the very act of hand-raising must indicate some motivation to progress, otherwise students would disengage (see \textit{negative emotion during debugging struggle}). Therefore, the drivers at play here may be a lack of motivation to problem-solve independently of the teacher, or attitudes of low self-efficacy.

Several \textit{domain-specific} drivers for debugging may also contribute to the hands-up problem. For example, students learning a block- or text-based programming language may have misconceptions (see \textit{conceptual difficulties} and \citep{MisconceptionsBeginnerProgrammer}) that make errors difficult to spot without the help of a more knowledgeable other. Teachers in our study also discussed drivers specific to learning a TBPL, such as the \textit{difficulty of getting syntax exactly right} and \textit{difficult to interpret PEMs}. Again, previous work has repeatedly identified these as difficulties with debugging in a TBPL \citep{UnderstandingSyntaxBarrier, AllSyntaxErrorsNotEqual, ErrorMessageWorkingGroup}. We suggest how these could be alleviated in Section \ref{sec:counters}.

Generally, domain-specific drivers can be viewed through cognitive load theory \citep{OriginalCognitiveLoadArticle, CognitiveArchitecture20YearsLater}. If a student's working memory is near-saturated when programming, errors and difficulty with problem-solving follow \citep{OriginalCognitiveLoadArticle}. Hand-raising is thus a logical behaviour, as students lack the cognitive capacity to debug themselves. However, if many students are experiencing a high load on working memory, the hands-up problem quickly follows.

\subsubsection{The Consequences}
Most teachers in our study alluded to many students raising their hands for debugging support, particularly when first learning to program. This is why we call it the hands-up \textit{problem}; in typical classroom settings, teachers simply cannot go around and effectively support every student \citep{ExpandingCSChallengesExperiences, CurrentPerspectivesDebugging}. This has both \textit{student-side} and \textit{teacher-side} consequences.

The \textit{student-side consequences} are multi-faceted. Once students raise their hands, teachers in our study explained how unlikely it was for students to progress until they came to help. Given the drivers discussed, this is not surprising, as students may not have the working memory to continue their debugging efforts. However, this means some students will linger in a state of being stuck or emotional angst for an extended period. Even when a teacher does come to provide support, the teacher-student interaction will likely be rushed if there is a time pressure or other students with their hands up \citep{CurrentPerspectivesDebugging, TeacherSupportForDebuggingPhysicalComputing}. Such a classroom atmosphere makes it very difficult for teachers to meaningfully guide students through the errors they are struggling with.

The emotional consequences of debugging are also important to consider. Feelings of frustration, anxiety, and disappointment are documented in this study (\textit{the emotional nature of learning to debug}) and previous work \citep{ProgrammingAssignmentsEmotionalToll, EDAProgrammingEmotions, ReflectionsSustainedDebuggingSupport}. If these feelings occur repeatedly, students may simply disengage in future encounters with errors (see \textit{negative emotion during debugging struggle}). If experiences of emotional arousal and lack of success are students' first with learning to program, they can negatively form students' programming self-efficacy \citep{OriginalSelfEfficacyArticle, ProgrammingSelfEfficacySelfAssessment}. Such experiences are therefore important to limit, but hard to do so if students are heavily reliant on teacher support.

There are also \textit{teacher-side consequences} of the hands-up problem. Some teachers described their programming lessons as `chaotic' and `busy', in part due to the number of hands raised. Some also reported feeling a duty to help every student requesting assistance. Failing to do so evoked feelings of inadequacy or disappointment (see \textit{the toll on the teacher of this reliance}). Furthermore, if teachers are regularly assessed or inspected, a sea of raised hands may give the impression of an unproductive lesson. This can again evoke negative feelings in teachers, who may overcompensate by providing too much assistance to students. As \citeauthor{LackPerserveranceAutonomyDebugging} states in their reflections on their in-class debugging practice, \textit{``in the midst of a busy classroom where as a teacher I was multitasking on many levels, it was all too easy to over-scaffold and provide too much direction, to give the appearance of a seemingly successful lesson''} \citep[p. 15]{LackPerserveranceAutonomyDebugging}.

\subsubsection{The Counters}\label{sec:counters}
So far, we have characterised the hands-up problem in fairly negative terms. However, it can be counteracted. Several teachers in our study report using teaching approaches that encourage teacher-independent debugging (see \textit{motivating teacher-independent debugging} and \citet{CombattingLearnedHelplessness, CurrentPerspectivesDebugging, TeacherDiagnosticInterventionSkills}). We refer to these as \textit{counters} to the hands-up problem. Like the drivers, the counters can be divided into \textit{domain-agnostic} and \textit{domain-specific}.

\textit{Domain-agnostic} counters are strategies that teachers of any subject can apply to alleviate the hands-up problem. For instance, teachers may refuse to help students until they evidence some teacher-independent problem-solving (see \textit{refusing to help students} and \citet{CombattingLearnedHelplessness, CurrentPerspectivesDebugging}). Some may also encourage collaboration with peers before providing support (see \textit{encouraging helpful collaboration}).

There are also a variety of \textit{domain-specific} counters that support students with the debugging process. Teachers reported referring students to appropriate resources and tools (see \textit{signposting to resources for assistance} and \citet{SupportingStudentsScienceInquiry, ReflectionsSustainedDebuggingSupport, TeacherSupportForDebuggingPhysicalComputing}) and modelling the debugging process to the whole class (see \textit{guiding students through debugging strategies} and \citet{ReflectionsSustainedDebuggingSupport}). Some also taught heuristics for `decoding' PEMs due to their often-ambiguous nature (see \textit{utilising programming error messages}). An obvious benefit of these counters is that the whole class can benefit from them simultaneously. An attitudinal counter mentioned by some teachers is the enforcement of a positive error culture (see \textit{promoting a positive error culture} and \citet{TowardDebuggingPedagogy}). 

As well as the domain-specific counters that teachers mentioned, debugging-specific teaching approaches developed in prior work could be useful counters. Systematic debugging processes (e.g., \citep{CarverImprovingChildrensDebugging, SystematicProcessMichaeli}) or debugging-specific tooling (e.g., \citep{PythonTutor, Ladebug, SystematicDebuggingLogicalErrors}) may provide students with effective debugging strategies to employ when encountering errors. Additionally, features of the programming environment students use could alleviate some of the drivers specific to TBPLs. For example, frame-based environments allow students to write text-based programs while reducing the aforementioned syntax barrier \citep{FrameBasedEditing, StrypePoster2024}, and more informative PEMs (e.g., \citep{EffectiveApproachErrorMessageEnhancement, ErrorMessageReadability, HighPrecisionFeedbackSyntaxErrors}) would help reduce the dependency on the teacher for feedback. While the curriculum taught by teachers in our study mandates the teaching of a TBPL, secondary school teachers could instead teach a BBPL as a counter in itself, though this may come at the cost of perceived authenticity \citep{ToBlockOrNotToBlock}.

These strategies may take time for students to become accustomed to \citep{CombattingLearnedHelplessness}. However, the teachers who report consistently using these counters also report their students becoming more empowered to attempt teacher-independent debugging \citep{CurrentPerspectivesDebugging}. Ultimately, this gives the teacher more time to provide more meaningful support to a smaller number of struggling students.

Among the teachers we interviewed, those who reported employing counters tended to be more confident and had decades of experience. This was less true for the less confident teachers, leaving them more susceptible to the consequences of the hands-up problem. More generally, the prevalence of these counters in computing classrooms is unclear. Results from \citet{ComputingInCurriculumChallengesStrategies} suggest only a fraction of teachers have strategies for teaching debugging and resilience to errors, with a lack of time to teach programming probably a limiting factor.

\subsection{The Implications}
Although we only interviewed nine teachers, we expect our findings align with other teachers' experiences for two reasons. First, previous researchers and teachers have reported struggles in line with the hands-up problem \citep{CombattingLearnedHelplessness, ComputingInCurriculumChallengesStrategies, CurrentPerspectivesDebugging, LackPerserveranceAutonomyDebugging, StumpTheTeacher}. Second, the commonality of non-specialist computing teachers and reported lack of content knowledge by some \citep{ExpandingCSChallengesExperiences, ComputingInCurriculumChallengesStrategies} hints at a lack of debugging PCK necessary for effectively supporting students with debugging.

For teachers who report counters to the hands-up problem, experience or professional development seemed to be the reason. This begs the question: how else can counters to the hands-up problem be developed? Or, in other words, how can less experienced or confident teachers teach debugging more confidently? We did not identify any themes related to this in our analysis and thus only provide initial recommendations for future research.

One suggestion is through professional development. Some studies have already reported successful debugging-specific PD, which have been participatory in nature \citep{DoINeedToKnowIfTheTeacher, EffectsOfProfessionalDevelopmentK12Debugging, SupportingStudentsScienceInquiry}. However, given that some teachers still report findings aligned with the hands-up problem, debugging-specific PD is either not widespread enough or not being effectively translated into practice. If the former is the case, debugging-related PD needs to be developed or expanded. This does not need to solely focus on debugging; more emphasis on debugging could be included in programming PD workshops, given the intertwined nature of these skills.

Similarly, informing teachers about pedagogical approaches for debugging developed by prior work has huge potential. Although long-term evidence for many debugging-specific teaching approaches is lacking \citep{DebuggingInterventionLitReview}, the increasing amount of research in this area at least provides teachers with some suggestions. Some teachers would likely adapt these to their classroom practice \citep{EffectsOfProfessionalDevelopmentK12Debugging}, which could provide further insight into the efficacy and applications of these approaches. How teachers can be informed about debugging-specific approaches in a way that encourages classroom adoption is an important research-to-practice consideration, but outside the scope of this article.

These suggestions relate to the improvement of teachers' `debugging PCK'. Despite the challenges debugging presents in programming classrooms, we are not aware of a conceptualisation of debugging PCK in the research literature. A debugging PCK would inevitably have some commonalities with conceptualisations of programming PCK \citep{ProgrammingPCK}, such as the motivation for teaching it and common student difficulties. It would also include debugging-specific pedagogy, techniques for effectively managing teacher-student interactions, and other aforementioned counters to the hands-up problem. Therefore, even if a teacher does not have time to explicitly teach the debugging process, a more developed debugging PCK would equip them to deal with the hands-up problem. Once the concept of a debugging PCK is more well-developed, it can be spread through professional development.

\section{Conclusions and Further Work}
This paper has reported secondary school teachers' experiences with the teaching and learning of debugging in their programming lessons, with a specific focus on text-based programming. Through thematically analysed semi-structured interviews, we report teachers' strategies for supporting students with debugging, their observations of how students engage with the debugging process, and some challenges they face related to debugging.

A particular trend among our themes was the common student reliance on the teacher when encountering and struggling to resolve errors. This corroborates with previous reports of computing teachers' classroom experiences \citep{CombattingLearnedHelplessness, ComputingInCurriculumChallengesStrategies, CurrentPerspectivesDebugging, LackPerserveranceAutonomyDebugging, StumpTheTeacher}. We conceptualised our results through the \textit{hands-up problem}, the in-class difficulty deriving from the common student tendency to raise their hand when struggling to complete a task. There are several \textit{drivers} for the frequent hand-raising that contributes to the hands-up problem, which have generally negative \textit{consequences}. Students are often left stuck and frustrated, and teachers flustered. Usually through experience, some teachers developed \textit{counters} to the hands-up problem. These consisted of general classroom management techniques and debugging-specific teaching approaches. Over time, these counters help students to become less reliant on the teacher and more resilient to errors in their programs.

This paper presents the hands-up problem as an initial conceptualisation of a significant challenge teachers face in programming lessons as a result of student struggle with debugging. This conceptualisation should be developed through further studies of teachers' experiences with teaching programming in the classroom. Where previous studies have used interviews \citep{CurrentPerspectivesDebugging} and simulations of teacher-student interactions \citep{DoINeedToKnowIfTheTeacher, SupportingStudentsScienceInquiry, TeacherDiagnosticInterventionSkills}, methods such as classroom observations would help to paint a more well-rounded picture of teachers' lived experiences. This work is increasingly important to conduct given the increasing number of school curricula mandating the teaching of text-based programming (e.g., \citep{UKSecondaryComputingCurriculum, BritishColumbiaADSTCurriculum, PolandCSCurriculumChallengesSolutions, ProgrammingSkillsInLatinAmerica, Swedish2024Curriculum}).

Developing a model for debugging PCK is another area of further work we recommend. This would be a useful way of synthesising research into beginner debugging education at a school level and beyond. A debugging PCK should include approaches developed by both researchers and teachers, which could be iteratively tested and refined in the classroom. This could then inform further work on debugging-specific professional development and pedagogy.

Finally, further investigation into pedagogical approaches that facilitate positive experiences with learning to debug should be conducted. This is already happening, but future research should investigate these approaches on a larger scale over longer periods of time to better understand their effectiveness \citep{DebuggingInterventionLitReview}. Supporting teachers' debugging pedagogies will help to cultivate classroom environments where errors are confidently handled rather than feared.

\begin{acknowledgement}[title={Acknowledgments}]
We are very grateful to all of the teachers we interviewed for sharing their invaluable experiences. Without their participation, this research would not have been possible.

The first author gives all praise to the Lord Jesus Christ for blessing him with the wisdom and opportunity to conduct this research.
\end{acknowledgement}

\bibliographystyle{infedu}
\bibliography{bibliography.bib, study-repository}


\end{document}